\documentclass[conference]{IEEEtran}
\usepackage{amsmath,amssymb,amsfonts}
\usepackage{algorithm,algorithmic}
\usepackage{graphicx}
\usepackage{amsthm}

\usepackage{mathtools}

\begin{document}

{
    \theoremstyle{plain}
    \newtheorem{example}{Example}
}    

{
    \theoremstyle{plain}
    \newtheorem{theorem}{Theorem}
}    

\renewcommand{\algorithmicrequire}{\textbf{Input:}}
 \renewcommand{\algorithmicensure}{\textbf{Output:}}
%
\title{Extended target Poisson multi-Bernoulli mixture trackers based on sets of trajectories}
%
%
%

\author{\IEEEauthorblockN{Yuxuan Xia, Karl Granstr\"{o}m\\Lennart Svensson}
\IEEEauthorblockA{\textit{Dept. of Electrical Eng.} \\
\textit{Chalmers Univ. of Tech.}\\
Gothenburg, Sweden \\
firstname.lastname@chalmers.se}
\and
\IEEEauthorblockN{{\'A}ngel F. Garc{\'\i}a-Fern{\'a}ndez}
\IEEEauthorblockA{\textit{Dept. of Electrical Eng. and Electronics} \\
\textit{Univ. of Liverpool}\\
Liverpool, United Kingdom \\
angel.garcia-fernandez@liverpool.ac.uk}
\and
\IEEEauthorblockN{Jason L. Williams}
\IEEEauthorblockA{\textit{CSIRO Data61}\\
Brisbane, Australia \\
jason.williams@data61.csiro.au}
}

\maketitle

\begin{abstract}
The Poisson multi-Bernoulli mixture (PMBM) is a multi-target distribution for which the prediction and update are closed. By applying the random finite set (RFS) framework to multi-target tracking with sets of trajectories as the variable of interest,
the PMBM trackers can efficiently estimate the set of target trajectories. This paper derives two trajectory RFS filters for extended target tracking, called extended target PMBM trackers. Compared to the extended target PMBM filter based on sets on targets, explicit track continuity between time steps is provided in the extended target PMBM trackers.
\end{abstract}


%
\IEEEpeerreviewmaketitle

\section{Introduction}
Multi-target tracking (MTT) denotes the process of estimating the set of target trajectories based on a sequence of noise-corrupted measurements, including missed detections and false alarms \cite{mtt}. Conventional MTT algorithms are usually tailored to the “point target” assumption: each target is modeled as a point without spatial extent, and each target gives rise to at most one measurement per time scan. However, modern high-resolution radar and lidar sensors make the “point tar- get” assumption unrealistic, because with such sensors it is common that a target gives rise to multiple measurements per time scan. The tracking of such a target leads to the extended target tracking problem, where the objective is to recursively determine the extent and kinematic states of the targets over time. A detailed overview of extended target tracking literature is given in \cite{extendedoverview}.

The focus of this paper is on extended targets. A target may give rise to more than one measurement if multiple resolution cells of the sensor are occupied by a single target. A common extended target measurement model is the inhomogeneous Poisson Point Process (PPP) \cite{gilholm2005poisson}. At each time step, a Poisson distributed random number of measurements are generated, spatially distributed around the target. For tracking multiple extended targets, random finite sets (RFSs) \cite{rfs} can be used to model the problem \cite{rfsextended}. The framework of RFSs was developed to provide a systematic methodology for dealing with MTT problems involving time-varying number of targets, where targets and measurements are modelled as random sets. The PPP extended target model has been integrated into several computationally feasible RFSs-based filters, see, e.g., \cite{mahler2009phd,phdextended2,phdextended3,cphdextended,lmbextended,pmbmextended,pmbmextended2,xia2018extended}.

In the RFSs formulation, the multi-target filtering density contains the information of the target states at the current time step. Exact closed-form (and computationally tractable) solutions to the RFSs-based multi-target Bayes filter are based on multi-target conjugate priors. Multi-target conjugate prior was defined in \cite{glmbconjugateprior} as meaning that ``\textit{if we start with the proposed conjugate initial prior, then all subsequent predicted and posterior distributions have the same form as the initial prior.}''

Two well-established MTT conjugate priors found in the literature are the Poisson multi-Bernoulli mixture (PMBM) \cite{pmbmpoint}, based on unlabelled RFSs\footnote{Labels can be incorporated into the PMBM density, but only in an ad hoc manner.}, and the Delta generalized labelled multi-Bernoulli ($\delta$-GLMB) \cite{glmbconjugateprior}, based on labelled RFSs. The PMBM conjugate prior consists of a PPP representing targets which are hypothesized to exist but have never been detected, and a multi-Bernoulli mixture (MBM) representing targets that have been detected at some point in time. For both the PMBM and $\delta$-GLMB multi-target densities, conjugacy has been shown for both point targets \cite{pmbmpoint,glmbconjugateprior} and extended targets \cite{pmbmextended2,lmbextended}.

The relations between the two point target conjugate priors are explored in \cite{pmbmpoint2}, where it is shown that the PMBM density has a more efficient structure than the $\delta$-GLMB density, with fewer global hypotheses. Simulation studies have shown that filters based on the PMBM conjugate prior in general outperform the filters based on the $\delta$-GLMB conjugate prior, in terms of filtering performance and computational cost, see \cite{performanceevaluation} for point target and \cite{pmbmextended2} for extended target. However, PMBM filters (without labels) seemingly do not provide explicit track continuity between time steps.

One approach to address the lack of track continuity is to add unique labels to the target states and estimate target states from the multi-target filtering density \cite{glmbconjugateprior,garcia2013two,aoki2016labeling}. The $\delta$-GLMB filter \cite{glmbconjugateprior,lmbextended} is a labelled filter when the birth model is a labelled multi-Bernoulli RFS. Labelling works well in many cases but it becomes problematic in challenging situations, e.g., when target birth is independent and identically distributed, or when targets get in close proximity and then separate; this can lead to problems with switching, see \cite{granstrom2018poisson} for an example of this. 

An appealing approach to ensuring track continuity for RFSs-based multi-target filters is to generalize the concept of RFSs of targets to RFSs of trajectories. A formulation of the target tracking problem as RFSs of trajectories was provided in \cite{svensson2014target,garcia2016multiple}. Within this set of trajectories framework, the goal of MTT is to recursively compute the posterior density over the set of trajectories, which contains full information about the target trajectories. From a trajectory and a given data association hypothesis, we can infer at all times the location of the target. Thus, there is no need to label targets upon initialization.

Closed-form PMBM filtering recursions for point targets, based on the set of trajectories framework, have been derived in \cite{granstrom2018poisson}. This enables us to leverage on the benefits of the PMBM recursions, while also obtaining track continuity. It is therefore of interest to show that the trajectory PMBM filtering recursions are also closed for extended target tracking.

In this work, we present prediction and update equations of two trajectory PMBM filters for extended target tracking: one in which the set of current (i.e., ``alive'') trajectories is tracked, and one in which the set of all trajectories (both ``dead'' and ``alive'') up to the current time is tracked. We call these tracking algorithms extended target PMBM trackers, to distinguish them from the extended target PMBM filter \cite{pmbmextended2}, which is for sets of target states. We also present results from a simulation study where we compare the tracking results to the $\delta$-GLMB filter \cite{lmbextended}, in terms of trajectory estimation error.

The paper is organized as follows. In Section II, we introduce the modeling assumption and background on set of trajectories. In Section III, we present prediction and update equations for the two extended target PMBM trackers. An implementation of the proposed tracking algorithms is given in Section IV. Simulation results are presented in Section V, and conclusions are drawn in Section VI.

\section{Background}
In this section, we first outline the modeling assumptions utilized in this work. Next, we give a brief introduction to RFSs of trajectories. Then, we introduce the generalized transition and measurement model in the framework of set of trajectories.

\subsection{Modelling assumptions}
In the traditional RFSs of targets problem formulation, target states and measurements are represented in the form of finite sets \cite{rfs}. Let $x_k$ denote a target state at time $k$, and let $z_k$ denote a measurement at time $k$. The set of measurements obtained at time step $k$ is denoted as $\mathbf{z}_k$. We utilize the standard multi-target dynamic model and the standard extended target measurement model, defined in the following.

\subsubsection{Standard multi-target transition model} 
New targets appear in the surveillance area independently of any existing targets. Targets arrive at each time according to a non-homogeneous Poisson RFS with birth intensity $D^b_k(x_k)$. Given a target state $x_k$, the target survives with a
probability $P^S(x_k)$ and moves with a single target transition density $\pi(x_k|x_{k-1})$.

\subsubsection{Standard extended target measurement model}
The set of measurements $\mathbf{z}_k$ is a union of a set of clutter measurements and a set of target-generated measurements; the sets are assumed to be independent. The clutter is modeled as a Poisson RFS with Poisson rate $\lambda^{\mathrm{FA}}$ and spatial distribution $c(z)$, and the clutter Poisson intensity is $\kappa(z) = \lambda^{\mathrm{FA}} c(z)$. Each extended target is detected with probability $P^D(x_k)$. If the extended target is detected, the target-generated measurements are modeled as a Poisson RFS with Poisson rate $\gamma(x_k)$ and spatial distribution $\phi(z_k|x_k)$.

The conditional extended target measurement likelihood for a nonempty set of measurements $\mathbf{w}_k$ is the product of the target detection probability $P^D(x_k)$ and the Poisson density of target-generated measurements $\mathbf{w}_k$ \cite{pmbmextended2},
\begin{equation}
    \ell_{\mathbf{w}_k}(x_k) = P^D(x_k)e^{-\gamma(x_k)}\prod_{z_k\in \mathbf{w}_k}\gamma(x_k)\phi(z_k|x_k).
    \label{eq:measurementlikelihood}
\end{equation}
The effective detection probability for an extended target with state $x_k$ is the product of target detection probability $P^D(x_k)$ and the Poisson probability that the target generates at least one measurement $1-e^{-\gamma(x_k)}$. Accordingly, the probability that the target is not detected, or equivalently the conditional likelihood for an empty set of measurements, is
\begin{equation}
    \ell_{\emptyset}(x_k) = 1 - P^D(x_k) + P^D(x_k)e^{-\gamma(x_k)}.
\end{equation}

\subsection{Random finite sets of trajectories}


Let $\mathcal{X}$ represent the single target state space, e.g., $\mathcal{X}=\mathbb{R}^4$ if the state represents position and velocity in two dimensions. We use the trajectory state model presented in \cite{svensson2014target}, in which the trajectory state is a tuple $X = (\beta,\epsilon,x_{\beta:\epsilon})$, where $\beta$ is the discrete time of the trajectory birth, i.e., the time the trajectory begins; $\epsilon$ is the discrete time of the trajectory's end time. If $k$ is the current time, $\epsilon=k$ means that the trajectory is alive; $x_{\beta:\epsilon}$ is, given $\beta$ and $\epsilon$, the sequence of states
\begin{equation}
    x_{\beta},x_{\beta+1},...,x_{\epsilon-1},x_{\epsilon},
\end{equation}
where $x_k\in\mathcal{X}$ for all $k\in\{\beta,...,\epsilon\}$. This gives a trajectory of length $l=\epsilon - \beta + 1$ time steps. The trajectory state space at time $k$ is \cite{garcia2016multiple}
\begin{equation}
    \mathcal{T}_k = \uplus_{(\beta,\epsilon)\in I_k}\{\beta\}\times\{\epsilon\}\times\mathcal{X}^{\epsilon-\beta+1},
\end{equation}
where $\uplus$ denotes disjoint set union, $I_k=\{(\beta,\epsilon):0\leq\beta\leq\epsilon\leq k\}$ and $\mathcal{X}^l$ denotes the Cartesian products of $\mathcal{X}$. The trajectory state density of $X_k$ given measurements up to and including time $k^{\prime}\leq k$ factorizes as follows
\begin{equation}
    p_{k|k^{\prime}}(X) = p_{k|k^{\prime}}(x_{\beta:\epsilon}|\beta,\epsilon)P_{k|k^{\prime}}(\beta,\epsilon),
    \label{eq:tradensity}
\end{equation}
where, if $\epsilon < \beta$, then $P_{k|k^{\prime}}(\beta,\epsilon)$ is zero. Integration for single trajectory densities is performed as follows \cite{garcia2016multiple},
\begin{multline}
    \int p(X)dX =\\\sum_{\beta,\epsilon}\left[\int...\int p(x_{\beta},...,x_{\epsilon}|\beta,\epsilon)dx_{\beta}...dx_{\epsilon}\right]P(\beta,\epsilon).
\end{multline}
A set of trajectories is denoted as $\mathbf{X}_k\in\mathcal{F}(\mathcal{T}_k)$, where $\mathcal{F}(\mathcal{T}_k)$ is the set of all finite subsets of $\mathcal{T}_k$. Let $g(\mathbf{X}_k)$ be a real-valued function on a set of trajectories, then the set integral is
\begin{multline}
    \int g(\mathbf{X}_k)\delta\mathbf{X}_k \triangleq \\ g(\emptyset) + \sum_{n=1}^{\infty}\frac{1}{n!}\int\cdots\int g(\{X^1_k,\ldots,X^n_k\})dX^1_k\cdots dX_k^n.
\end{multline}

Two basic building blocks of RFSs-based MTT are the Poisson RFS and the Bernoulli RFS. A trajectory Poisson RFS has density
\begin{equation}
    f^{\text{ppp}}(\mathbf{X}) = e^{-\int D(X^{\prime})dX^{\prime}}\prod_{X\in\mathbf{X}}D(X),
    \label{eq:ppp}
\end{equation}
where the trajectory Poisson RFS intensity $D(\cdot)$ is defined on the trajectory state space $\mathcal{T}_k$, i.e., realizations of the Poisson RFS are trajectories with a birth time, a time of the most recent state, and a state sequence. 

A trajectory Bernoulli RFS has density
\begin{equation}
    f^{\text{ber}}(\mathbf{X}) = \begin{cases}
        1-r,& \mathbf{X}=\emptyset\\
        rf(X),& \mathbf{X}=\{X\}\\
        0,& \text{otherwise}
    \end{cases}
    \label{eq:bernoulli}
\end{equation}
where $f(\cdot)$ is a single trajectory density, cf. (\ref{eq:tradensity}), and $r$ is the Bernoulli probability of existence. Together, $f(\cdot)$ and $r$ can be used to find the probability that the target trajectory existed at a specific time, or find the probability that the target state was in a certain area at a certain time. Trajectory multi-Bernoulli RFS and trajectory MBM RFS are both defined analogously to target multi-Bernoulli RFS and target MBM RFS: a trajectory multi-Bernoulli is the disjoint union of a multiple trajectory Bernoulli RFS; trajectory MBM RFS is an RFS whose density is a mixture of trajectory multi-Bernoulli densities.

\subsection{Transition models for sets of trajectories}
In the standard multi-target transition model, target birth at time $k$ is modelled by a Poisson RFS. We write the birth intensity as
\begin{subequations}
    \begin{align}
        D^B_k(X)&=D^{B,x}_k(x_{\beta:\epsilon}|\beta,\epsilon)\Delta_k(\epsilon)\Delta_k(\beta),\\
        D^{B,x}_k(x_{k:k}|k,k) &= D^b_k(x_k),
    \end{align}
    \label{eq:pppbirth}%
\end{subequations}
where $\Delta(\cdot)$ denotes Kronecker delta function. 

As in \cite{granstrom2018poisson}, we focus on two different MTT formulations: the set of current trajectories, where the objective is to estimate the trajectories of targets that are still present in the surveillance area at the current time; and the set of all trajectories, where the objective is to estimate the trajectories for all targets that have been present at any times. For the set of current trajectories, $\mathbf{X}_k$ is the set of trajectories for which $0\leq \beta \leq \epsilon = k$. For the set of all trajectories, $\mathbf{X}_k$ is the set of trajectories for which $0\leq\beta\leq\epsilon\leq k$. The probability of survival as a function on trajectories at time $k$ is defined as
\begin{equation}
    P^S_k(X) = P^S(x_{\epsilon})\Delta_k(\epsilon).
\end{equation}
The transition density for the trajectories depends on the problem formulation.

\subsubsection{Transition model for the set of current trajectories}
The Bernoulli RFS transition density without birth is %
\begin{subequations}
    \begin{align}
        \begin{split}
        {}&f^c_{k|k-1}(\mathbf{X}|\mathbf{X}^{\prime}) =\\ &\begin{cases}
            1,& \mathbf{X}^{\prime}=\emptyset,\mathbf{X}=\emptyset\\
            1-P^S_{k-1}(X^{\prime}),& \mathbf{X}^{\prime}=\{X^{\prime}\},\mathbf{X}=\emptyset\\
            P^S_{k-1}(X^{\prime})\pi^c(X|X^{\prime}),& \mathbf{X}^{\prime}=\{X^{\prime}\},\mathbf{X}=\{X\}\\
            0,& \text{otherwise}
        \end{cases}
        \end{split}\\
        &\pi^c(X|X^{\prime}) = \pi^{c,x}(x_{\beta:\epsilon}|\beta,\epsilon,X^{\prime})\Delta_{\epsilon^{\prime}+1}(\epsilon)\Delta_{\beta^{\prime}}(\beta),\\
        &\pi^{c,x}(x_{\beta:\epsilon}|\beta,\epsilon,X^{\prime}) = \pi^x(x_{\epsilon}|x^{\prime}_{{\epsilon^{\prime}}})\delta_{x^\prime_{\beta^{\prime}:\epsilon^{\prime}}}(x_{\beta:\epsilon-1}),
    \end{align}
    \label{eq:transition_current}%
\end{subequations}%
where $\delta(\cdot)$ denotes Dirac delta function. In this model, if the target disappears, or ``dies'', then the entire trajectory will no longer be a member of the set of current trajectories. If the trajectory survives, then the trajectory is extended by one time step.
\subsubsection{Transition model for the set of all trajectories}
The Bernoulli RFS transition density without birth is
\begin{subequations}
    \begin{align}
        \begin{split}
        {}&f^a_{k|k-1}(\mathbf{X}|\mathbf{X}^{\prime}) =\\ &\begin{cases}
            1,& \mathbf{X}^{\prime}=\emptyset,\mathbf{X}=\emptyset\\
            \pi^a(X|X^{\prime}),& \mathbf{X}^{\prime}=\{X^{\prime}\},\mathbf{X}=\{X\}\\
            0,& \text{otherwise}
        \end{cases}
        \end{split}\\
        &\pi^a(X|X^{\prime}) = \pi^{a,x}(x_{\beta:\epsilon}|\beta,\epsilon,X^{\prime})\pi^{\epsilon}(\epsilon|\beta,X^{\prime})\Delta_{\beta^{\prime}}(\beta),\\
        {}&\pi^{\epsilon}(\epsilon|\beta,X^{\prime}) = \begin{cases}
            1,&\epsilon = \epsilon^{\prime}<k-1\\
            1-P^S_{k-1}(X^{\prime}), &\epsilon = \epsilon^{\prime}=k-1\\
            P^S_{k-1}(X^{\prime}),& \epsilon = \epsilon^{\prime}+1=k\\
            0,& \text{otherwise}
            \end{cases}\\
        \begin{split}
        {}&\pi^{a,x}(x_{\beta:\epsilon}|\beta,\epsilon,X^{\prime}) = \\ &\begin{cases}
            \delta_{x^{\prime}_{\beta^{\prime}:\epsilon^{\prime}}}(x_{\beta:\epsilon}),&\epsilon=\epsilon^{\prime}\\
            \pi^x(x_{\epsilon}|x^{\prime}_{\epsilon^{\prime}})\delta_{x^{\prime}_{\beta^{\prime}:\epsilon^{\prime}}}(x_{\beta:\epsilon-1}).&\epsilon=\epsilon^{\prime}+1
        \end{cases}
        \end{split}
    \end{align}
    \label{eq:transition_all}%
\end{subequations}
\noindent In this model, the interpretation of the probability of survival is that it governs whether or not the trajectory ends, or if it extends by one more time step. However, importantly, regardless of whether or not the trajectory ends, the trajectory remains in the set of all trajectories.

For both transition models, the predicted set of trajectories is the union of the birth process and the set of trajectories that arise from previous set of trajectories.

\subsection{Single trajectory measurement model}
The standard extended target measurement model is extended by defining a Bernoulli measurement density as follows:
\begin{subequations}
    \begin{align}
        \begin{split}
        {}&\varphi_k(\mathbf{w}_k|\mathbf{X}) =\\ &\begin{cases}
            1,& \mathbf{X}=\emptyset,\mathbf{w}_k=\emptyset\\
            \ell_{\emptyset}(X),& \mathbf{X}=\{X\},\mathbf{w}_k=\emptyset\\
            \ell_{\mathbf{w}_k}(X),& \mathbf{X}=\{X\},\mathbf{w}_k\neq \emptyset\\
            0,& \text{otherwise}
        \end{cases}
        \end{split}\\
        &\ell_{\mathbf{w}_k}(X)=\ell_{\mathbf{w}_k}(x_{\epsilon})\Delta_k(\epsilon), \label{eq:measurementmodel_trajLik}
    \end{align}
    \label{eq:measurementmodel}%
\end{subequations}
where \eqref{eq:measurementmodel_trajLik} means that the target can only cause detections if it is present at the current time, $\epsilon = k$. Note that the notation in \eqref{eq:measurementmodel_trajLik} is abused since $\ell_{\mathbf{w}_k}(x_{\epsilon})$ is undefined for $\epsilon\neq k$.

\section{Extended target PMBM trackers}
We extends the closed-form filtering recursion of the PMBM tracker for point targets \cite{pmbmpoint,granstrom2018poisson} to extended targets. The trajectory PMBM has density
\begin{subequations}
    \begin{align}
        f_{k|k^{\prime}}(\mathbf{X}_k) &= \sum_{\mathbf{X}_k^u\uplus\mathbf{X}_k^d=\mathbf{X}}f_{k|k^{\prime}}^{\textrm{ppp}}(\mathbf{X}_k^u)\sum_{a\in\mathcal{A}_{k|k^{\prime}}}w^a_{k|k^{\prime}}f_{k|k^{\prime}}^a(\mathbf{X}_k^d),\\
        f_{k|k^{\prime}}^{\textrm{ppp}}(\mathbf{X}^u_k) &= e^{-\int D_{k|k^{\prime}}^u(X^{\prime})dX^{\prime}}\prod_{X\in\mathbf{X}_k^u}D_{k|k^{\prime}}^u(X),\\
        f_{k|k^{\prime}}^a(\mathbf{X}_k^d) &= \sum_{\uplus_{i^{\prime}\in\mathbb{T}_{k|k^{\prime}}}\mathbf{X}_k^{i^\prime}=\mathbf{X}_k^d}\prod_{i\in\mathbb{T}_{k|k^{\prime}}}f_{k|k^{\prime}}^{i,a^i}(\mathbf{X}^i_k),
        \label{eq:mbmdensity}
    \end{align}
    \label{eq:pmbm}%
\end{subequations}
where the set of trajectories $\mathbf{X}_k$ is an independent union of a Poisson RFS $\mathbf{X}^u_k$ with intensity $D^u_{k|k^{\prime}}$ and an MBM RFS $\mathbf{X}_k^d$ with Bernoulli parameters $r^{i,a^i}_{k|k^{\prime}}$ and $f^{i,a^i}_{k|k^{\prime}}(\cdot)$, cf. (\ref{eq:bernoulli}). 

The Poisson RFS represents trajectories that are hypothesized to exist, but have never been detected, i.e., no measurement has been associated to them. In the MBM in (\ref{eq:mbmdensity}), $\mathbb{T}_{k|k^{\prime}}$ is a track table with $n_{k|k^{\prime}}$ tracks, $a\in\mathcal{A}_{k|k^{\prime}}$ is a global data association hypothesis, and for each global hypothesis $a$ and for each track
$i\in\mathbb{T}_{k|k^{\prime}}$, $a^i$ indicates which local track hypothesis is used in the global hypothesis. Each global hypothesis is a collection of single trajectory hypothesis, one from each track. For each track, there are $h^i_{k|k^{\prime}}$ single trajectory hypotheses. 

Let $m_k$ be the number of measurements at time $k$ and $j\in\mathbb{M}_k=\{1,...,m_k\}$ be an index to each measurement. Let $\mathcal{M}^{k}$ denote a set of tuples $(\tau,j)$, where $\tau\leq k$ and $j\in \mathbb{M}_{\tau}$. Let $\mathcal{M}^k(i,a^i)\subseteq \mathcal{M}^{k}$ denote the history of measurements that are associated to track $i$ in hypothesis $a^i$. Compared to point target models \cite{pmbmpoint}, here $\mathcal{M}^k(i,a^i)$ can contain more than one element that corresponds to the same time step, see below for a simple example. 
\begin{example}
If $\mathcal{M}^5(i,a^i) = \{(3,1),(3,2),(5,8)\}$, then $a^i$ hypothesizes that the $i$-th hypothesized target was first detected at time $3$ by measurements 1 and 2, at time 4 a missed detection occurred, and at time 5 it was detected by measurement 8.
\end{example}
\noindent For each global hypothesis $a=(a^1,...,a^{n_{k|k^{\prime}}})$, it satisfies that
\begin{subequations}
    \begin{align}
 \bigcup_{i\in\mathbb{T}_{k|k^{\prime}}}\mathcal{M}^k(i,a^i) &= \mathcal{M}^{k},  \\
         \mathcal{M}^k(i,a^i) \cap \mathcal{M}^k(i^{\prime},a^{i^{\prime}}) &= \emptyset~\forall~i\neq i^{\prime}.
    \end{align}
\end{subequations}

In the following, we will show how the trajectory PMBM density is predicted and updated, in order to track either the set of current trajectories, or the set of all trajectories. Analogous to the point target PMBM tracks \cite{granstrom2018poisson}, the two different extended target PMBM trackers, based on two different problem formulations, have the same update step but different prediction steps. For compactness, we denote the inner product of two functions $h(\cdot)$ and $g(\cdot)$, as $\langle h;g\rangle = \int h(x)g(x)dx$. 

\subsection{Prediction step}
The prediction steps for the set of current trajectories and the set of all trajectories are, respectively, given in the two theorems below.
\begin{theorem}
    Assume that the set of current trajectories distribution from the previous time step $f_{k-1|k-1}(\mathbf{X}_{k-1})$ is given by (\ref{eq:pmbm}), that the transition model is (\ref{eq:transition_current}), and that the birth model is a trajectory Poisson RFS with intensity of the form (\ref{eq:pppbirth}). Then the predicted distribution for the next step $f_{k|k-1}(\mathbf{X}_{k})$ is given PMBM, cf. (\ref{eq:pmbm}), with:
    \begin{subequations}
        \begin{align}
            D^u_{k|k-1}(X_k) &= D^B_k(X_k) + \left\langle D^u_{k-1|k-1};\pi^cP^S_{k-1} \right\rangle,\\
            n^i_{k|k-1} &= n^i_{k-1|k-1},\\
            h^i_{k|k-1} &= h^i_{k-1|k-1},\\
            w^{i,a^i}_{k|k-1}&=w^{i,a^i}_{k-1|k-1}~\forall~a^i,\\
            r^{i,a^i}_{k|k-1} &= r^{i,a^i}_{k-1|k-1}\left\langle f^{i,a^i}_{k-1|k-1};P^S_{k-1}\right\rangle~\forall~a^i,\\
            f^{i,a^i}_{k|k-1}(X_k) &= \frac{\left\langle f^{i,a^i}_{k-1|k-1};\pi^cP^S_{k-1}\right\rangle }{\left\langle f^{i,a^i}_{k-1|k-1};P^S_{k-1}\right\rangle}~\forall~a^i.
        \end{align}
        \label{eq:predictioncurrent}
    \end{subequations}
    \label{theorem1}
\end{theorem}
\begin{theorem}
    Assume that the set of all trajectories distribution from the previous time step $f_{k-1|k-1}(\mathbf{X}_{k-1})$ is given by (\ref{eq:pmbm}), that the transition model is (\ref{eq:transition_all}), and that the birth model is a trajectory Poisson RFS with intensity of the form (\ref{eq:pppbirth}). Then the predicted distribution for the next step $f_{k|k-1}(\mathbf{X}_{k})$ is given PMBM, cf. (\ref{eq:pmbm}), with:
    \begin{subequations}
        \begin{align}
        D^u_{k|k-1}(X_k) &= D^B_k(X_k) + \left\langle D^u_{k-1|k-1};\pi^a \right\rangle,\\
            n^i_{k|k-1} &= n^i_{k-1|k-1},\\
            h^i_{k|k-1} &= h^i_{k-1|k-1},\\
            w^{i,a^i}_{k|k-1}&=w^{i,a^i}_{k-1|k-1}~\forall~a^i,\\
            r^{i,a^i}_{k|k-1} &= r^{i,a^i}_{k-1|k-1}~\forall~a^i,\\
            f^{i,a^i}_{k|k-1}(X_k) &= \left\langle f^{i,a^i}_{k-1|k-1};\pi^a \right\rangle~\forall~a^i.
            \label{eq:pmbm_alltra_prediction}
        \end{align}
        \label{eq:predictionall}
    \end{subequations}
    \label{theorem2}
\end{theorem}

\subsection{Update step}
We present ``\emph{track-oriented}'' (TO) extended target PMBM trackers, where a track is initiated for each non-empty subset of the measurement set $\mathbf{z}_k$ at each time $k$; this is analogous to how, in point target tracking, a track is initiated for each measurement at each time $k$. 

The update step of the extended target PMBM trackers is presented in the following theorem.  We denote the power set of $\mathbf{z}_k$, i.e., the set of all subsets of $\mathbf{z}_k$, as $\mathcal{P}(\mathbf{z}_k)$. Further, we denote the $p$th nonempty element in $\mathcal{P}(\mathbf{z}_k)$ as $\mathbf{w}^p_k$ ($p\in\{1,...,|\mathcal{P}(\mathbf{z}_k)|-1\}$) by ordering the elements of $\mathcal{P}(\mathbf{z}_k)$ in an arbitrary manner; the set of measurement indices of $\mathbf{w}^p_k$ is denoted as $\{j_1,...,j_{|\mathbf{w}^p_k|}\}$.

\begin{theorem}
    Assume that the predicted distribution $f_{k|k-1}(\mathbf{X}_k)$ is given by (\ref{eq:pmbm}), that the single trajectory measurement model is (\ref{eq:measurementmodel}), and that the clutter is a Poisson RFS with intensity $\kappa(z)$. Then, the updated distribution $f_{k|k}(\mathbf{X}_k)$ (updated with the measurement set $\mathbf{z}_k$) is a PMBM, cf. (\ref{eq:pmbm}), with $n_{k|k} = n_{k|k-1} + |\mathcal{P}(\mathbf{z}_k)|-1$, and
    \begin{equation}
        D^u_{k|k}(X_k) = \ell_{\emptyset}(X_k)D^u_{k|k-1}(X_k).
        \label{eq:poissonupdate}
    \end{equation}
    
    For tracks continuing from previous time steps ($i\in\{1,...,n_{k_k-1}\}$), a hypothesis is included for each combination of a hypothesis from a previous time, and either a missed detection or an update using a nonempty subset of $\mathbf{z}_k$. The number of hypotheses becomes $h^i_{k|k}=|\mathcal{P}(\mathbf{z}_k)| h^i_{k|k-1}$\footnote{\label{ft1}Aspects regarding practical implementation and computational tractability will be discussed in next section.}. 
    
    For missed detection hypotheses ($i\in\{1,...,n_{k|k-1}\}$, $a^i\in\{1,...,h_{k|k-1}\}$):
    \begin{subequations}
        \begin{align}
            \mathcal{M}^k(i,a^i) &= \mathcal{M}^{k-1}(i,a^i),\\
            w^{i,a^i}_{k|k} &=w^{i,a^i}_{k|k-1}\left(1-r^{i,a^i}_{k|k-1}+r^{i,a^i}_{k|k-1} \left\langle  f^{i,a^i}_{k|k-1};\ell_{\emptyset}\right\rangle \right),\\
            r^{i,a^i}_{k|k} &=\frac{r^{i,a^i}_{k|k-1}\left\langle f^{i,a^i}_{k|k-1};\ell_{\emptyset} \right\rangle}{1-r^{i,a^i}_{k|k-1}+r^{i,a^i}_{k|k-1} \left\langle  f^{i,a^i}_{k|k-1};\ell_{\emptyset}\right\rangle},\\
            f^{i,a^i}_{k|k}(X_k) &= \frac{ \ell_{\emptyset}(X_k)f^{i,a^i}_{k|k-1}(X_k) }{\left\langle f^{i,a^i}_{k|k-1};\ell_{\emptyset} \right\rangle}.
        \end{align}
        \label{eq:misseddetection}
    \end{subequations}
    
    For hypotheses updating existing tracks ($i\in\{1,...,n_{k|k-1}\}$, $\tilde{a}^i\in\{1,...,h^i_{k|k-1}\}, p\in\{1,...,|\mathcal{P}(\mathbf{z}_k)|-1\}$, i.e., the previous hypothesis $\tilde{a}^i$, updated with nonempty measurement set $\mathbf{w}^p_k$):\footnote{A hypothesis at previous time with $r^{i,a^i}_{k|k-1}=0$ needs not to be updated since the corresponding posterior weight would be zero. For simplicity, the hypothesis numbering does not account for this exclusion.}
    \begin{subequations}
        \begin{align}
            a^i &= \tilde{a}^i + h^i_{k|k-1}p \\
            \mathcal{M}^k(i,a^i) &= \{(k,j_1),\dots,(k,j_{|\mathbf{w}^p_k|})\}\cup\mathcal{M}^{k-1}(i,\tilde{a}^i),\\
            w^{i,a^i}_{k|k} &=w^{i,a^i}_{k|k-1}r^{i,\tilde{a}^i}_{k|k-1}\left\langle f^{i,\tilde{a}^i}_{k|k-1};\ell_{\mathbf{w}^p_k} \right\rangle,\\
            r^{i,a^i}_{k|k} &= 1,\\
            f^{i,a^i}_{k|k}(X_k) &= \frac{\ell_{\mathbf{w}^p_k}(X_k)f^{i,\tilde{a}^i}_{k|k-1}(X_k)}{\left\langle f^{i,\tilde{a}^i}_{k|k-1};\ell_{\mathbf{w}^p_k}\right\rangle}.
        \end{align}
        \label{eq:updateexisting}
    \end{subequations}
    Finally, for new tracks, ($i\in\{n_{k|k-1}+p\}$, $p\in\{1,...,|\mathcal{P}(\mathbf{z}_k)|-1\}$, i.e., the new track commencing on measurement set $\mathbf{w}^p_k$)\footnotemark[2]:
    \begin{subequations}
        \begin{align}
            h^i_{k|k} &= 2,\\
            \mathcal{M}^k(i,1) &= \emptyset,~w^{i,1}_{k|k} = 1,~r^{i,1}_{k|k}=0,\\
            \mathcal{M}^k(i,2) &= \{(k,j_1),\dots,(k,j_{|\mathbf{w}^p_k|})\},\\
            w^{i,2}_{k|k} &= \begin{cases}
        \kappa(\mathbf{w}^p_k)+\left\langle D^u_{k|k-1};\ell_{\mathbf{w}^p_k}\right \rangle,& |\mathbf{w}^p_k|=1\\
        \left\langle D^u_{k|k-1};\ell_{\mathbf{w}^p_k}\right \rangle,& |\mathbf{w}^p_k| > 1,
    \end{cases}\\
            r^{i,2}_{k|k} &= \begin{cases}
        \frac{\left\langle D^u_{k|k-1};\ell_{\mathbf{w}^p_k} \right\rangle}{\kappa(\mathbf{w}^p_k)+\left\langle D^u_{k|k-1};\ell_{\mathbf{w}^p_k}\right \rangle},& |\mathbf{w}^p_k|=1\\
        1,& |\mathbf{w}^p_k| > 1,
    \end{cases}\\
    f^{i,2}_{k|k}(X_k) &= \frac{\ell_{\mathbf{w}^p_k}(X_k)D^u_{k|k-1}(X_k)}{\left\langle D^u_{k|k-1};\ell_{\mathbf{w}^p_k} \right\rangle }.
        \end{align}%
        \label{eq:newtracks}%
    \end{subequations}%
    \label{theorem3}%
\end{theorem}\hfill$\square$

Note that from the last part of the theorem, the probability that $\mathbf{w}^p_k$ is clutter if it has only one element is incorporated into the existence probability $r$.


\begin{figure*}[!t]
    \centering
    \includegraphics[width=\textwidth]{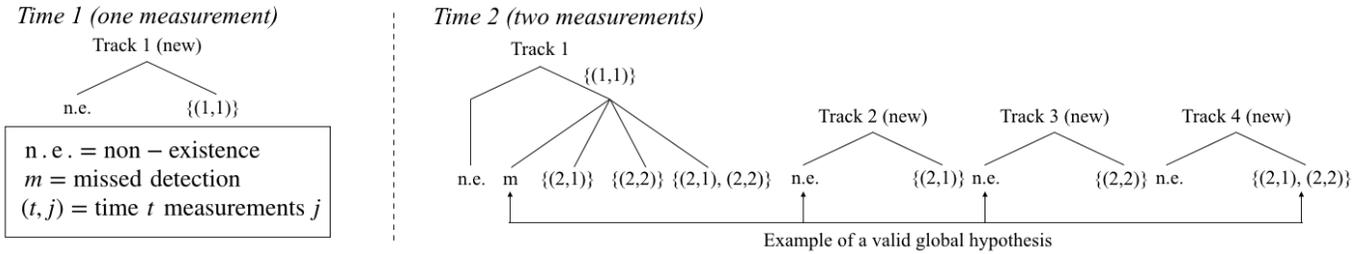}
    \caption{Tracks and hypotheses maintained by the extended target PMBM trackers. Structure after time 1 is shown at left (assuming there was one measurement). Structure after time 2 is shown at right (assuming there were two measurements). A new track is created for each nonempty subset of the measurements received. New tracks each contain two hypotheses; one hypothesizing that at least one of the measurements goes with a previously existing track and hence the new track is not required, and the other hypothesizing that the measurements go with the new track, capturing both the possibility that it is the result of target detected for first time, or a false alarm (if the cardinality of the measurement subset is one). Each track from prior distribution is continued, incorporating a hypothesis for each prior hypothesis corresponding to missed detection, and for each combination of prior hypothesis and nonempty subset of new measurements. Non-existence hypotheses are continued without branching. The line with arrows at the bottom shows an example of a global hypothesis, i.e., a choice of one single trajectory hypothesis from the tree for each track, in which measurement at every time is used exactly once.}
    \label{fig:hypothesistree}
\end{figure*}

\subsection{Discussion}
A global hypothesis can be considered as a partitioning of all measurements received so far into subsets, where each subset is hypothesized to correspond to a particular potential target\footnote{The term ``potential target'' is used because single trajectory hypotheses correspond to Bernoulli distributions.}. Each single trajectory hypothesis explains the association of
each measurement received so far that are hypothesized to correspond to the same potential target. The weight of global hypothesis $a$ is $w^a_{k|k^{\prime}}\propto\prod_{i\in\mathbb{T}_{k|k^{\prime}}}w^{i,a^i}_{k|k^{\prime}}$, where $w^{i,a^i}_{k|k^{\prime}}$ is the weight of single trajectory hypothesis $a^i$ from track $i$. 

The single trajectory density and the intensity of the Poisson RFS is a mixture density of the form
\begin{equation}
    p(X) = \sum_{t}w^{t}p^{t}(x_{\beta:\epsilon}|\beta,\epsilon)\Delta_{e^{t}}(\epsilon)\Delta_{b^{t}}(\beta),
    \label{eq:densityrepresentation}
\end{equation}
where each mixture component is characterized by a weight $w^{t}$, a distinct birth time $b^{t}$, a distinct most recent time $e^{t}$ where $b^{t}\leq e^{t}$ for all ${t}$\footnote{Neither the birth time $\beta$ nor the most recent time $\epsilon$ is deterministic.}, and a state sequence density $p^{t}(\cdot)$. For the weights we have that $\sum_{t}w^{t}=1$ if $p(\cdot)$ is a density, and $\sum_{t}w^{t}\geq 0$ if $p(\cdot)$ is an intensity function, e.g., a Poisson RFS intensity. This type of state density facilitates simple representations for the state sequence $x_{\beta:\epsilon}$ (either the state of a trajectory that is still present, or the state of a dead trajectory), conditioned on $\beta$ and $\epsilon$.

Two different extended target PMBM trackers result from the theorems: a PMBM tracker for the set of current trajectories is given by the prediction in Theorem \ref{theorem1} and the update in Theorem \ref{theorem3}; a PMBM tracker for the set of all trajectories is given by the prediction in Theorem \ref{theorem2} and the update in Theorem \ref{theorem3}. Both PMBM trackers are TO. For each track there is a hypothesis tree, where each hypothesis corresponds to different data association sequences for the track. An example of this hypothesis structure is illustrated in Figure \ref{fig:hypothesistree}. The predictions (\ref{eq:predictioncurrent}) and (\ref{eq:predictionall}) preserve the number of tracks and hypotheses; however, the prediction (\ref{eq:pmbm_alltra_prediction}) results in additional mixture components in (\ref{eq:densityrepresentation}). In the update step, the number of global hypotheses increases rapidly due to the complexity of the data association problem in extended target tracking. Exact expressions for the number of possible data associations and number of global hypotheses under a Poisson birth model can be found in \cite[Section V]{granstrom2018likelihood} and \cite[Section V]{pmbmextended2}, respectively.


\section{Implementation}
In this section, we discuss reduction methods that can be used to keep the computational complexity of the extended target PMBM tracker at a tractable level. We also present pseudo code for the update and the prediction step of the (TO) extended target PMBM tracker.

\subsection{Handling the data association}
First, the number of single trajectory hypotheses to be created in the update step is reduced using gating. For extended target tracking, the gates should take into account both the position and the extent of the target, as well as state uncertainties. 


Second, the number of global hypotheses can be reduced by only considering data association events with high likelihoods. In extended target tracking, the data association problem is usually handled in two stages, see, e.g., \cite{pmbmextended2,lmbextended}. First, clustering methods are used to find a set of different ways in which the measurements can be clustered; second, assignment methods, e.g., Murty's algorithm \cite{murty}, are used to assign measurement clusters to targets, based on the assumption that only one measurement cluster can be assigned to a track. As an alternative to using clustering and assignment to find a subset of associations, random sampling methods \cite{granstrom2018likelihood,granstrom2017pedestrian,Fatemi2017gibbssamplingradarmapping} can be used, which work directly on maximizing the data association likelihood. Both simulation and experiment results have shown that the sampling methods have the advantage that they work equally well for both spatially close and well separated targets, see \cite{granstrom2018likelihood}. Hence, in the PMBM trackers we use sampling.

Third, after an updated PMBM density has been computed, global hypotheses whose updated weight fall below a threshold are pruned. Note that pruning does not affect the symmetry of the posterior. By doing so, we can prune single trajectory hypotheses (Bernoulli components) that are not included in the remaining global hypotheses. Further, we prune Bernoulli components with probability of existence smaller than a threshold. For the mixture representation of the Poisson RFS intensity, components with weights smaller than a threshold are pruned.

\subsection{Pseudo code for the update and the prediction}
In the TO implementation, global hypotheses are represented using a look-up table. The $(j,i)$th entry of the look-up table is the index of the single trajectory hypothesis in the $i$th track that is included in the $j$th global hypothesis. If the $j$th global hypothesis does not include any single trajectory hypothesis from the $i$th track, then the $(j,i)$th entry of the look-up table would be zero. This can either be the case of non-existence single trajectory hypothesis or the case that the single trajectory hypothesis is pruned due to its small existence probability.
\begin{example}
Let us consider the hypothesis structure illustrated in Figure \ref{fig:hypothesistree}. There are in total four valid global hypotheses. Assume that the single trajectory hypotheses in each track are indexed in left-to-right order, the maintained global hypotheses look-up table would be
$$ \begin{bmatrix}
1 & 0 & 0 & 1 \\
2 & 0 & 1 & 0 \\
3 & 1 & 0 & 0 \\
4 & 0 & 0 & 0 
\end{bmatrix},$$
where the first row corresponds to the global hypothesis example in Figure \ref{fig:hypothesistree}. Note that the corresponding entries of non-existence single trajectory hypotheses are zero since they are omitted in practical implementation.
\end{example}
The pseudo code for one update and prediction of the extended target PMBM tracker is given in Algorithm \ref{alg1}.
\begin{algorithm} 
\caption{Pseudo code for one prediction and update for extended target PMBM tracker} 
\label{alg1} 
\algsetup{
    linenosize=\footnotesize,
    linenodelimiter=.
 }
\begin{algorithmic}[1] 
    \REQUIRE Parameters of the PMBM posterior and the global hypotheses look-up table at the previous time step, and measurement set $\mathbf{z}$ at current time step.
    \ENSURE Parameters of the PMBM posterior and the global hypotheses look-up table at the current time step.
    \STATE Perform prediction, individually for each mixture component of Poisson intensity and each single trajectory hypothesis, see (\ref{eq:predictioncurrent}) or (\ref{eq:predictionall}).
    \FOR{$z \in \mathbf{z}$}
        \STATE Perform gating of $z$ w.r.t. each mixture component of Poisson intensity and each single trajectory density contained in Bernoulli RFSs.
    \ENDFOR
    \FOR{$a \in \mathcal{A}$ (rows of hypotheses look-up table)}
    \STATE For measurements that are inside the gate of existing targets, compute the subset of data associations.
    \STATE Based on the data association results, create new Bernoulli components for missed detection hypotheses (\ref{eq:misseddetection}), hypotheses updating existing tracks (\ref{eq:updateexisting}) and new tracks (\ref{eq:newtracks}).
    \STATE Update hypotheses look-up table.
    \ENDFOR
    \STATE For measurements that are not inside the gate of any existing targets but are inside the gate of at least one mixture component of Poisson intensity, create new tracks (\ref{eq:newtracks}) by clustering, and update hypotheses look-up table.
    \STATE Prune global hypotheses whose weight is below a threshold and update hypotheses look-up table.
    \STATE Prune Bernoulli components whose existence probability is below a threshold or do not appear in the truncated global hypotheses and update hypotheses look-up table.
    \STATE Merge duplicate global hypotheses and update hypotheses look-up table.
    \STATE Update the Poisson intensity, see (\ref{eq:poissonupdate}).
    \STATE Prune the mixture components in the Poisson intensity whose weight is below a threshold.
\end{algorithmic}
\end{algorithm}

\subsection{Single target model}
Solving the multiple extended target tracking problem requires not only an MTT framework, but also a single extended target model. There are several single extended target models available in the literature, see \cite{extendedoverview} for an overview. We chose to model the extended objects using the Gaussian inverse Wishart (GIW), or random matrix, model  \cite{koch2008bayesian,feldmann2011tracking}, in which the target shape is approximated by an ellipse. The GIW model was chosen because it is relatively simple to use and it has been integrated into many extended target tracking filters, including the PMBM filter \cite{pmbmextended} and the $\delta$-GLMB filter \cite{lmbextended}, making comparison easy. Further, smoothed kinematic and extent estimates can be obtained by performing GIW backwards filtering \cite{granstrom2019bayesian}. In \cite{cphdextended,phdextended}, the GIW model was extended to additionally estimate for each target the state dependent Poisson measurement rate $\gamma(x_k)$, resulting in the Gamma GIW (GGIW) model. Due to page limits, we refer the reader to \cite{pmbmextended,pmbmextended2} for GGIW-PMBM details.

\section{Simulation results}
In this section, we present Monte Carlo simulation results that compare the tracking performance of the GGIW-PMBM tracker for the set of all trajectories and the GGIW-GLMB filter \cite{lmbextended}. The filtering performance, i.e., how the estimated multi-target states compare to the true multi-target states at each time instant, of the PMBM filter \cite{pmbmextended2} and the $\delta$-GLMB filter has been evaluated in an exhaustive simulation study in \cite{pmbmextended2}. Here, we focus on the comparison of the tracking performance that fully account for errors between the estimated and the true set of tracks.

A two-dimensional Cartesian coordinate system is used to define measurement and target kinematic parameters. Targets follow a linear Gaussian nearly constant velocity model. The measurement model is also linear Gaussian. 

Trajectory estimation, or trajectory extraction, is the process of obtaining estimates of the set of trajectories (or set of targets) from the multi-target density. For GGIW-PMBM, an estimate of the set of trajectories is directly extracted from the from the MB component with the highest weight by taking the trajectory Bernoulli densities with existence probability larger than 0.5. For GGIW-GLMB, we first perform target state extraction at each time instant, see \cite{lmbextended} for details. Then target trajectories are formed by connecting target states with the same label.

For performance evaluation of extended target estimates with ellipsoidal extents, a comparison study in \cite{gwd} has shown that among several compared performance measures, the Gaussian Wassterstein Distance (GWD) metric is the best choice. To evaluate tracking performance, the trajectory metric $d(\cdot,\cdot)$ \cite{rahmathullah2016metric} was used by integrating the GWD as base distance measure, and with location/extent error cut-off $c=20$, order $p=1$, and track switch cost $\gamma=4$. In the simulated scenarios, we apply the metric at each time step, and normalize it by the time step. This allows a comparison of how the metric evolves over time in the scenario, as opposed to only computing the metric at the final time step.

\begin{figure}[!t]
    \centering
    \includegraphics[width=0.5\textwidth]{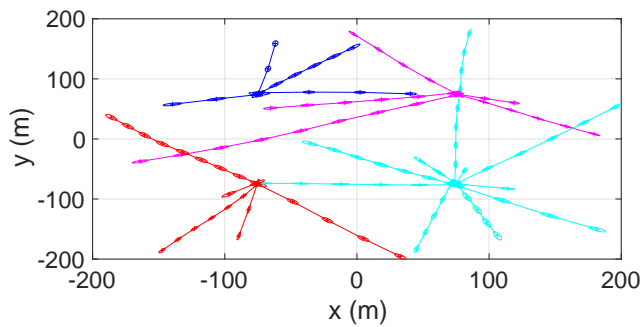}
    \caption{True target trajectories. Targets positions every 10 time steps are marked with a ''+`` sign, and their extents (3 sigma levels) with an ellipse. Targets born around the same position are marked with the same color.}
    \label{fig:groundtruth}
\end{figure}

We consider the scenario illustrated in Figure \ref{fig:groundtruth}, where 27 randomly generated targets were simulated for 100 time steps. The targets appear in, and disappear from, the surveillance area at different time steps. We choose target detection probability $P^D=0.9$, target measurement rate is randomly selected from $\{7,8,9\}$, target survival probability $P^S=0.99$, and uniformly distributed Poisson clutter with rate $\lambda^{\mathrm{FA}}=60$. The birth spatial density consists of four GGIW components, with positions in $[\pm75,\pm75]^{\mathrm{T}}$. For both GGIW-GLMB and GGIW-PMBM implementation, MB components with weight below $0.01$ are pruned. 

\begin{figure*}
    \centering
    \includegraphics[width=0.2\textwidth]{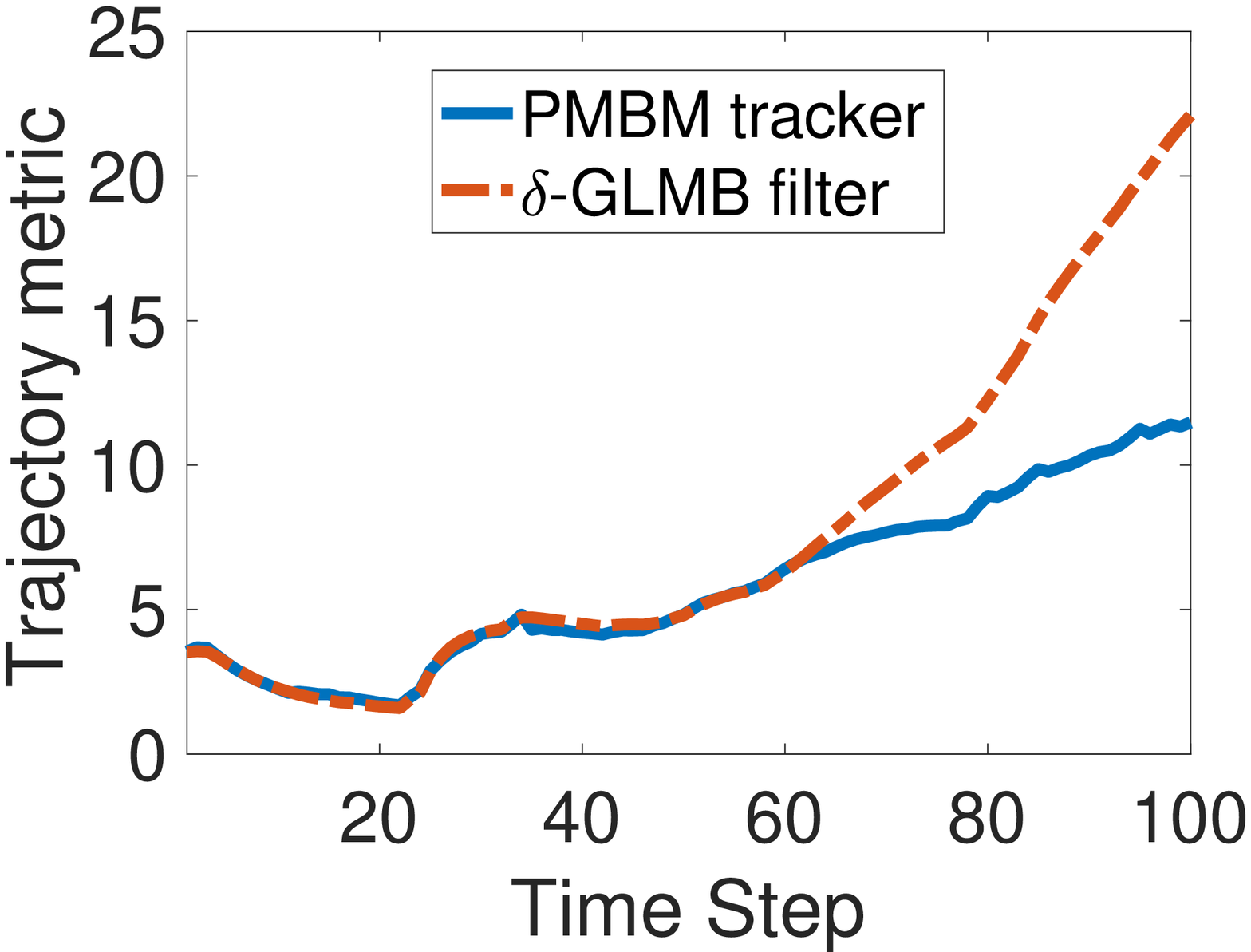}
    \hspace{-4mm}
    \includegraphics[width=0.2\textwidth]{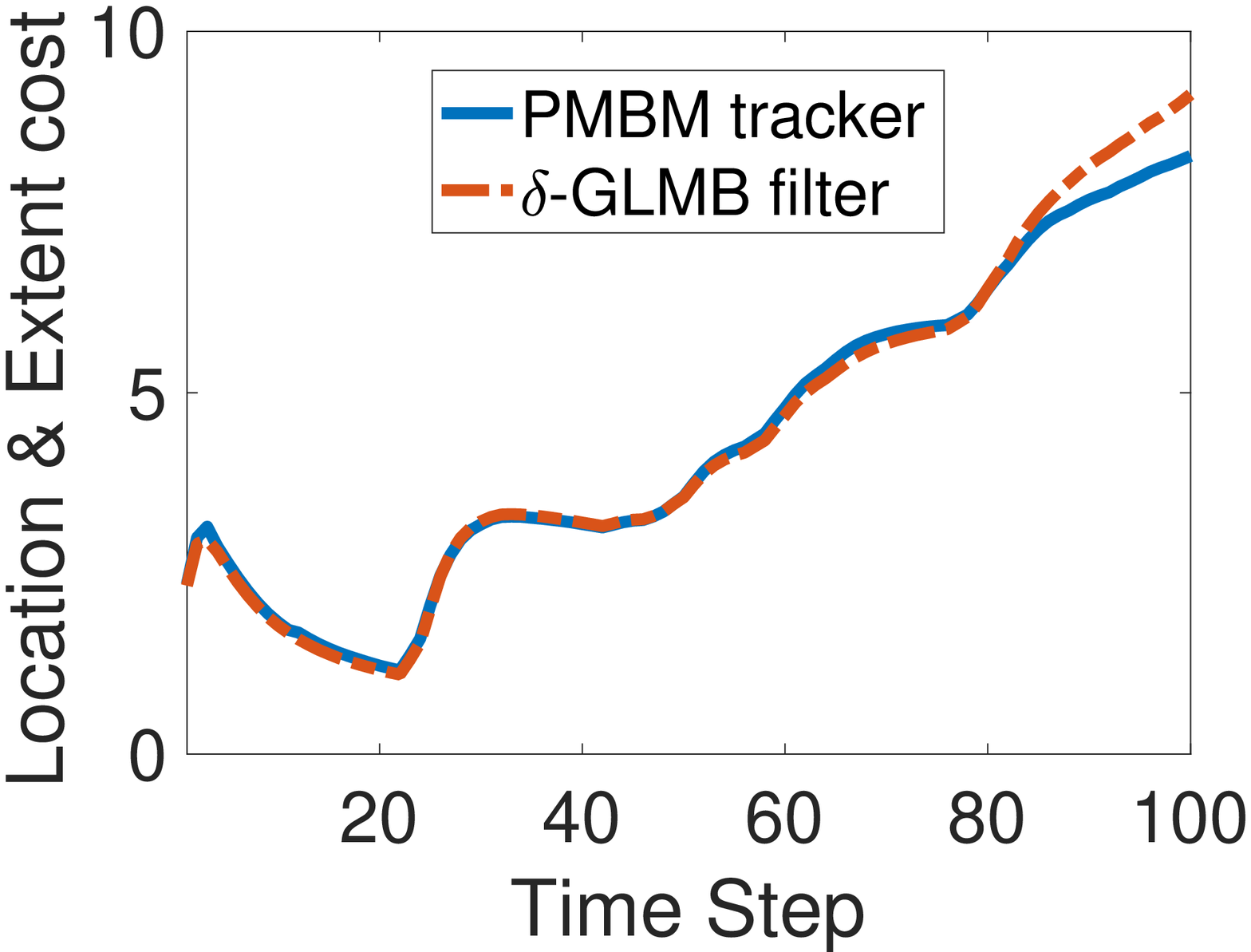}
    \hspace{-4mm}
    \includegraphics[width=0.2\textwidth]{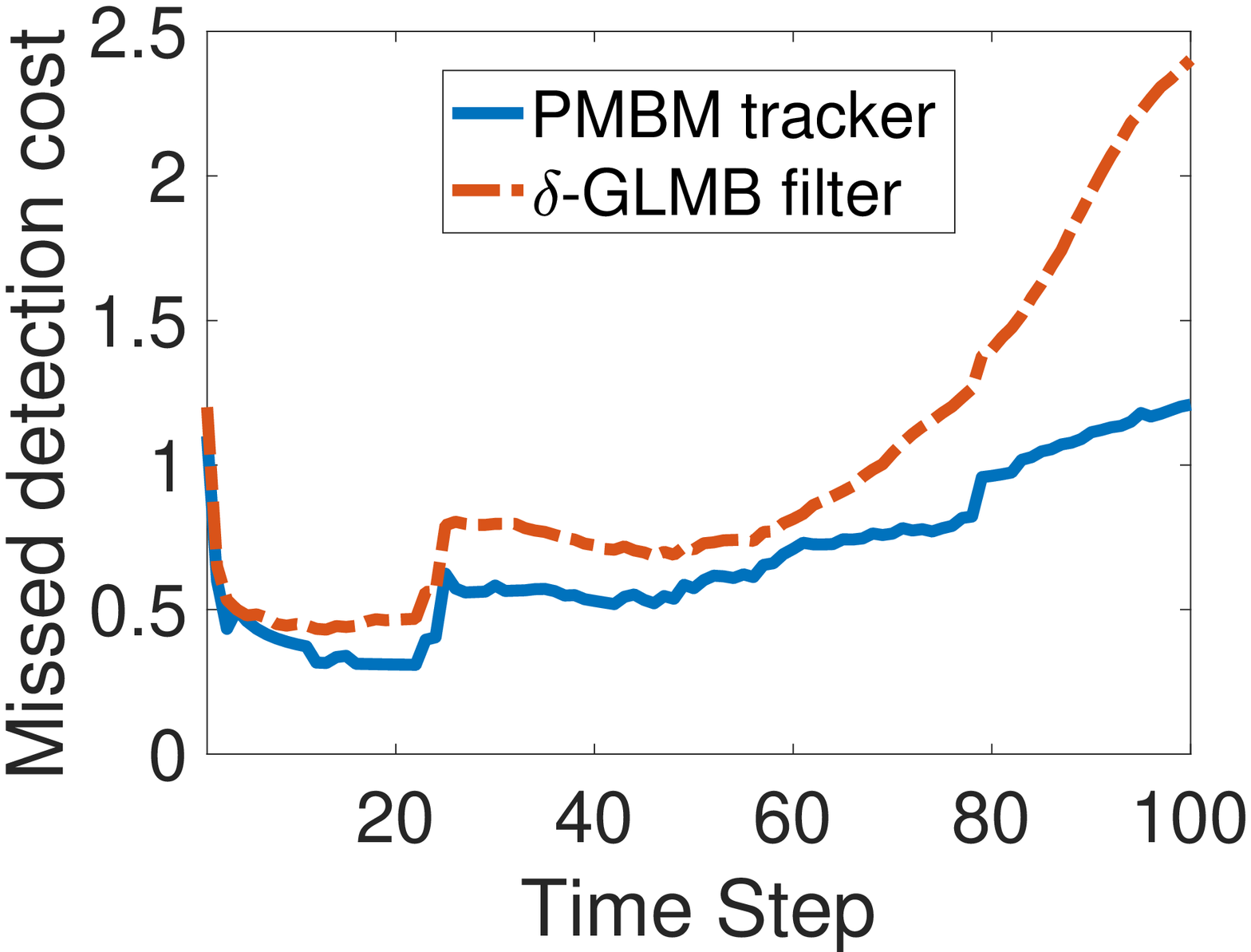}
    \hspace{-4mm}
    \includegraphics[width=0.2\textwidth]{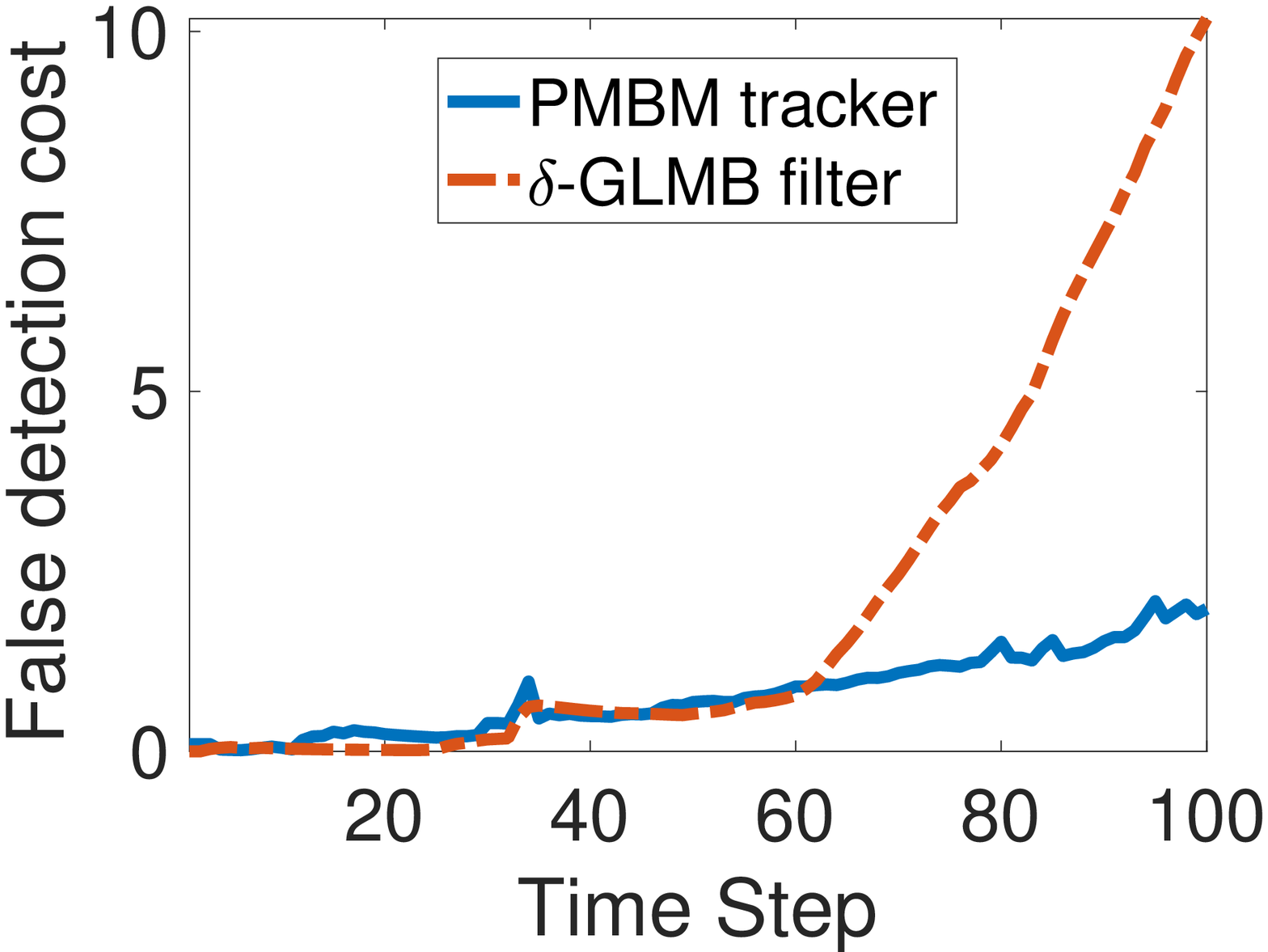}
    \hspace{-4mm}
    \includegraphics[width=0.2\textwidth]{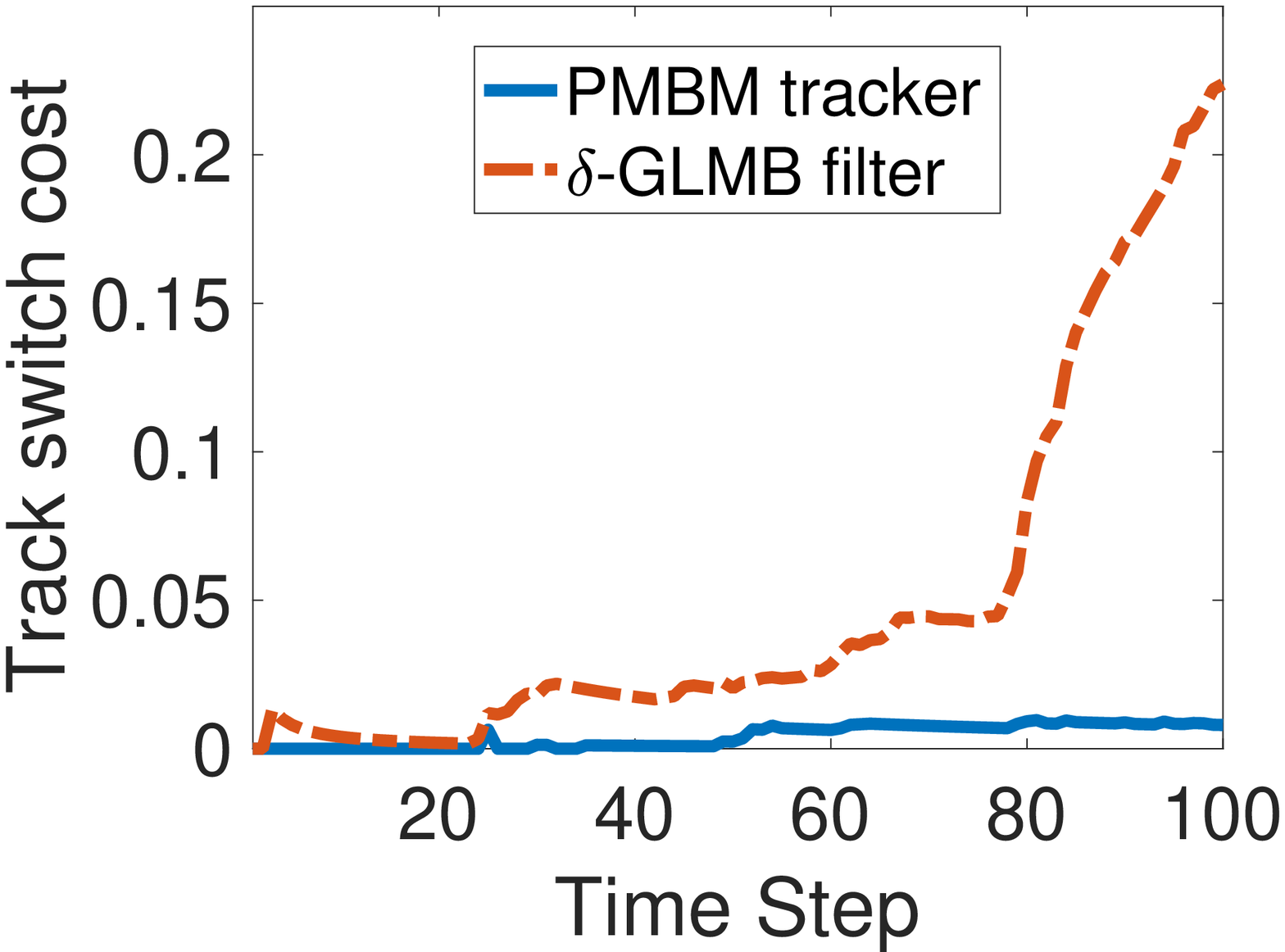}
    \caption{Results from simulated scenario with coalescence; the lines show mean over 100 Monte Carlo runs. The five subplots from left to right, respectively, present the total tracking error, target location and extent estimation error, missed detection error, false detection error and track switch error.}
    \label{fig:result}
\end{figure*}

The results in Figure \ref{fig:result} show the tracking errors averaged over 100 Monte Carlo runs for trajectory estimates extracted at each time step of the scenario. For this scenario, the average total time to process one full sequence of measurement sets was 1502s for GGIW-GLMB, and 45s for GGIW-PMBM\footnote{MATLAB implementation on 3 GHz Intel Core i5.}. We can see that the extended target PMBM tracker outperforms the extended target $\delta$-GLMB filter by a large margin in terms of both tracking error and computational complexity. The main reason is that PMBM is a more efficient parameterization than $\delta$-GLMB, which has deterministic existence probability \cite{pmbmpoint2}. In addition, it is noticeable that the extended target PMBM tracker presents negligible track switch error compared to the extended target $\delta$-GLMB filter. This is because that the extended target PMBM tracker always provide a valid trajectory, i.e., not one that is flipping between different hypotheses at different times \cite{granstrom2018poisson}.

\section{Conclusion}
In this paper we have presented two extended target PMBM trackers for the set of target trajectories that provide explicit track continuity between time steps. Future works include developing a smoothing-while-filtering GGIW implementation of the PMBM trackers and developing a multi-scan PMBM tracker that considers the multi-scan data association problem.


%





\ifCLASSOPTIONcaptionsoff
  \newpage
\fi



%


\bibliographystyle{IEEEtran}
\bibliography{mybibli.bib}

%








\end{document}